\documentclass[PAPER, cernpreprint=true, coverpage=false, atlasdraft=false, UKenglish, texmf, orcidlogo]{atlasdoc}
 
\usepackage{atlaspackage}
\usepackage{atlasbiblatex}
 
\usepackage{atlasphysics}

\graphicspath{{logos/}{figures/}}
 
\usepackage{mhchem}
\newcommand*{\xexe}{\ce{Xe}+\ce{Xe}\xspace}
\newcommand*{\Nrec}{\ensuremath{N_\mathrm{ch}^\mathrm{rec}}\xspace}
 
\newcommand*{\etaA}{\ensuremath{|\eta^a|}\xspace}

\AtlasTitle{Measurements of longitudinal flow decorrelations in $pp$ and Xe+Xe collisions with the ATLAS detector}
 
\AtlasVersion{1.5}

\AtlasAbstract{
Measurements of longitudinal flow decorrelations in 13~TeV $pp$ and 5.44~TeV Xe+Xe collisions with the ATLAS detector are presented.
The measurements are performed using the two-particle correlation method, combining charged-particle tracks within $|\eta| < 2.5$ with either calorimeter energy clusters or towers within $4.0<|\eta|<4.9$.
A template-based subtraction procedure is used to remove non-flow effects in both the $pp$ and the Xe+Xe analyses.
The dependence of the longitudinal flow decorrelations on the pseudorapidity separation between the particles is characterized via the slope parameter $F_n$ for the elliptic ($n=2$) and triangular ($n=3$) harmonic moments.
The results are reported as a function of charged-particle multiplicity for the $pp$ and Xe+Xe collision systems.
Comparing the data to a color string-based model of the initial geometry indicates that in $pp$ and peripheral Xe+Xe collisions, sub-nucleonic structure and fluctuations in longitudinal energy deposition are needed to describe the data.
}

\AtlasRefCode{HION-2022-17}
 
\PreprintIdNumber{CERN-EP-2023-184}

\AtlasJournal{Physical Review Letters}

\hypersetup{pdftitle={ATLAS document},pdfauthor={The ATLAS Collaboration}}
 
\begin{document}
 
\maketitle
 
Relativistic heavy-ion collisions are understood to create nucleus-sized droplets of quark-gluon plasma (QGP), where the time evolution is well described by nearly inviscid hydrodynamics~\cite{Heinz:2013th}.   This hydrodynamic modeling of collisions of large nuclei has been successfully  extended to collisions of smaller nuclei and even proton-proton (\pp) collisions~\cite{Nagle:2018nvi}.   An accurate description requires a quantitative description of the energy deposited at the earliest time of the collision.
Conversely, observables sensitive to hydrodynamic behavior also allow one to learn about the nuclear geometry, nucleon fluctuations, and even sub-nucleonic structure.
 
In the plane transverse to the beams, an anisotropic initial state (resulting from the intrinsic nuclear geometry and fluctuations at the nucleon, sub-nucleon or even gluon field level~\cite{Luzum:2013yya,Bernhard:2016tnd,Snyder:2020rdy}) evolves hydrodynamically to produce an azimuthally anisotropic final state with particle yields characterized by $dN/d\phi \propto 1 + 2\sum_{n=1}^\infty v_n \text{cos} [n(\phi-\Psi_n)]$, where the $v_n$ are Fourier moments referred to as flow coefficients, $\phi$ is the azimuthal angle in the transverse plane~\cite{coord}, and $\Psi_n$  is the $n^{\mathrm{th}}$ order event plane.
If the deposited energy is boost invariant~\cite{Bjorken:1982qr}, the $\Psi_{n}$ and $v_n$ are rapidity independent.
However, if the deposited energy is not boost invariant, i.e., not uniform along the beam direction, there may be a rapidity dependence of ~$\Psi_{n}$ and $v_n$, which is referred to as longitudinal flow decorrelation.   Numerous works have modeled longitudinal decorrelations and made specific predictions for experimental observables~\cite{Bozek:2010vz,Jia:2014ysa,Bozek:2015bna,Pang:2015zrq,Shen:2017bsr,Bozek:2017qir}. Other calculations, suggesting strong initial-state momentum anisotropies that may carry through to the final state, have a larger longitudinal decorrelation than ones based on longitudinally fluctuating energy deposition~\cite{Schenke:2022mjv,Schenke:2016ksl}.
 
Experiments have measured longitudinal decorrelations in Pb+Pb~\cite{CMS-HIN-14-012,HION-2016-04} and Xe+Xe~\cite{HION-2019-06} collisions.   Calculations based on initial-state color strings, longitudinally extended objects modeling long-range gluon fields~\cite{Sjostrand:2007gs}, such as \textsc{AMPT}~\cite{Lin:2004en}, yield a longitudinally dependent geometry from the finite length of the strings and
qualitatively describe these data~\cite{Pang:2015zrq}.
It is crucial to extend these measurements to \pp collisions and lower-multiplicity nucleus-nucleus collisions for understanding the role of sub-nucleonic degrees of freedom where their role is expected to dominate.

In this Letter, longitudinal decorrelations are measured in 13~\TeV \pp and 5.44~\TeV Xe+Xe collisions via
the comparison of large-rapidity-gap $(\Delta\eta\approx 7)$ to small-rapidity-gap $(\Delta\eta\approx 2)$ correlations, similar to the previous measurements, which construct correlations between charged-particle tracks around mid rapidity and a reconstructed event-plane angle $\Psi_{n}$ at very forward rapidity~\cite{CMS-HIN-14-012,HION-2016-04,HION-2019-06}.
In the case of \pp or low-multiplicity heavy-ion collisions, significant \textit{non-flow} correlations, which are not arising from many-body collective motion, such as from jets and particle decays, can mimic a longitudinal decorrelation signal, and must therefore be corrected for~\cite{Xu:2020koy}. It is also important to explore these effects in previously measured large systems~\cite{HION-2019-06}.
Here, the two-particle correlation (2PC) method is employed using charged-particle tracks and energy deposits in the forward calorimeter, which then allows the application of non-flow subtraction (NFS) procedures~\cite{HION-2015-09,HION-2013-04,CMS-QCD-10-002,CMS-HIN-16-010}.
Because the 2PC method is ubiquitous, a detailed understanding of the longitudinal dependence has great practical importance, for example, when comparing results between experiments with different $\eta$ acceptances~\cite{Nagle:2021rep,Zhao:2022ugy}.

The ATLAS detector~\cite{PERF-2007-01} is a general-purpose particle detector with a nearly full solid-angle acceptance and many detector subsystems, which makes it well suited for understanding the underlying correlations between particles across a large range in pseudorapidity.  The main subsystems used in this measurement are the tracking detectors, a subset of the calorimeter system, and the trigger system.
 
The inner detector (ID) provides charged-particle tracking within $|\eta|<2.5$ via a combination of subsystems:
a silicon-pixel detector (including the insertable B-layer~\cite{ATLAS-TDR-19,PIX-2018-001}), a silicon microstrip detector,
and a straw-tube transition-radiation tracker. 
 
The calorimeter provides full $\phi$ coverage within $|\eta|<4.9$ acceptance.
The forward regions, $3.2<|\eta|<4.9$, are instrumented with liquid-argon calorimeters (FCals) that include both electromagnetic (EM) and hadronic layers.
Sets of FCal cells at similar $\eta$-$\phi$ positions, over its multiple layers, are combined into \textit{towers}. Separately, FCal cells can be clustered in three dimensions into topologically connected clusters~\cite{PERF-2014-07}, referred to as topoclusters, which are more closely associated with single-particle energy depositions than the set of towers.
Topoclusters cannot be reconstructed in \xexe collisions due to the high occupancy in the FCal,  and hence topoclusters are only reconstructed in \pp data.
The energy in both the towers and clusters is evaluated at the electromagnetic scale~\cite{PERF-2011-03}.
 
A two-level trigger system~\cite{TRIG-2016-01} is used to select events.
The first trigger level is hardware-based and uses a subset of the detector information to restrict the accepted rate to be below \SI{100}{\kilo\hertz}.
This is followed by a software-based high-level trigger (HLT) stage, which reduces the accepted event rate to a few \SI{}{\kilo\hertz}, depending on the data-taking conditions.
An extensive software suite~\cite{ATL-SOFT-PUB-2021-001} is used in data simulation, in the reconstruction and analysis of real
and simulated data, in detector operations, and in the trigger and data acquisition systems of the experiment.

The 13~\TeV\ \pp analysis uses 1.7~pb$^{-1}$  of collision data recorded throughout 2015--2018 in which the average number of interactions per a crossing was less than $1.25$.
The collision data was recorded utilizing a variety of triggers, including minimum bias triggers and triggers enhancing the sample of high-multiplicity events~\cite{HION-2016-01}.
The $\sqrt{s_\mathrm{NN}}=5.44$~\TeV\ \xexe analysis uses 3~$\mu$b$^{-1}$ of collision data collected in 2017, with the triggers and offline criteria initially selecting minimum-bias events as described in Ref.~\cite{HION-2019-06}. \xexe events were further required to satisfy a reconstructed pseudo-rapidity gap requirement similar to that in Ref.~\cite{HION-2018-27} to reject the possible contribution of photo-nuclear processes to low-multiplicity events.
 
For both the \pp\ and \xexe data, the reconstructed primary vertex~\cite{ATL-PHYS-PUB-2015-026} is required to be within 100~mm of the center of the detector along the $z$-axis. In the \pp\ data, events with multiple collisions are suppressed by requiring only one reconstructed vertex.
Charged-particle tracks are reconstructed in the ID and must satisfy quality criteria related to the number of hits in different ID layers and their projected distance of closest approach to the primary vertex~\cite{STDM-2015-02}.
Tracks used in the correlation analysis are required to have $|\eta|<2.5$ and $0.3<\pt<5.0~\GeV$. However, the reconstructed charged-particle multiplicity, \Nrec, is defined as the number of tracks with $|\eta|<2.5$ and $\pt>0.4~\GeV$ in the event, to match the definition in previous measurements~\cite{HION-2019-06,HION-2018-23}.

The 2PC function measures the relative azimuthal angle between particles of type $a$ and so-called reference particles.  The $a$ particles are charged-particle tracks with $0.3<\pt<5.0~\GeV$, and the correlation function is evaluated as a function of their pseudorapidity, $\eta^a$.
For \pp collisions, the reference particles are topoclusters with $0.5< E_\text{T} <5.0$~\GeV\ and $4.0< \eta <4.9$, as measured in the FCal.
For \xexe collisions, FCal towers with $E_\text{T}<5~\GeV$ and $4.0<\eta<4.9$ are utilized.
As a component of the NFS in \xexe data, detailed below, 5.02~\TeV\ \pp collision data, recorded in 2017,
is used which utilizes FCal towers as well.
Due to the symmetric nature of the collisions and ATLAS acceptance, each event was also analyzed by swapping the pseudorapidity sign and using the other FCal.   The two results are compatible within uncertainties.
 
\begin{figure}
\centering
\includegraphics[width=0.49\textwidth]{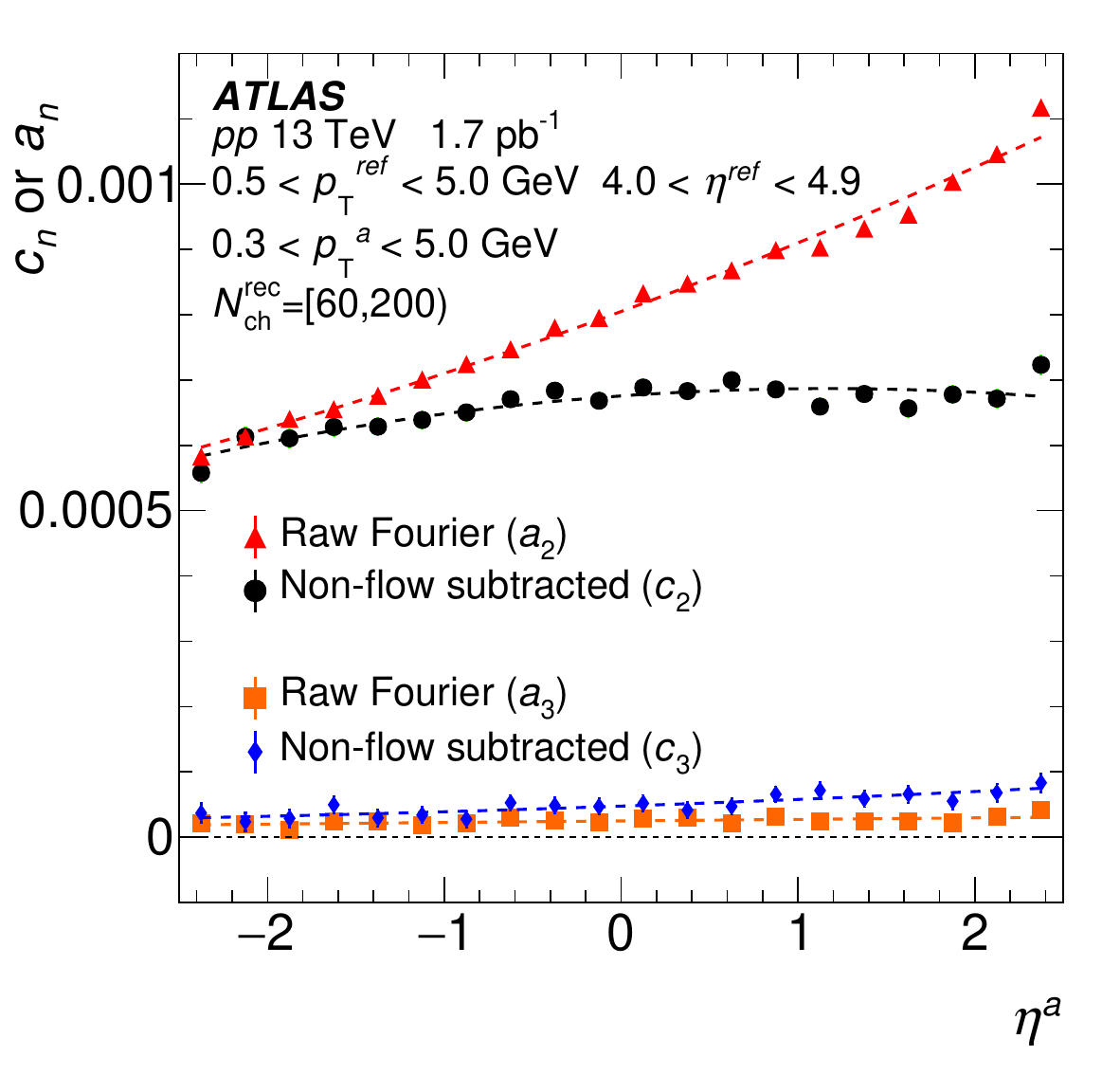}
\caption{Coefficients $a_n$ and $c_n$ in high-multiplicity 13~\TeV\ \pp events for $n = 2, 3$ as a function of track pseudorapidity $\eta^a$.  The $c_n$ values are obtained from the template fitting method.  The dashed lines represent a fit to a quadratic function in $\eta^a$ for each set of coefficients.
The pseudorapidity gap between the cluster and track pair is largest for negative $\eta^a$ and decreases moving to positive $\eta^a$. Only statistical uncertainties are shown. }
\label{fig:c2_exp}
\end{figure}
 
The track-topocluster or track-tower pair correlation functions of relative azimuthal angle, $\Delta\phi=\phi^{a}-\phi^\text{ref}$, were constructed for all possible pairs as a function of charged-particle track $\eta^a$. Each entry in the distributions was weighted by $1/\varepsilon_\text{trk}$, where $\varepsilon_\text{trk}$ is the $p_\mathrm{T}$- and $\eta$-dependent tracking efficiency.   The efficiency applied to the \pp (\xexe) data was obtained using events generated with \PYTHIA{8}~\cite{Sjostrand:2007gs} (\textsc{Hijing}~\cite{Wang:1991hta}) and passed through a \GEANT simulation~\cite{Agostinelli:2002hh,ATL-SOFT-PUB-2021-001} of the ATLAS detector.
For the track-tower pairs, there is an additional weighting by the calorimeter tower transverse energy to account for the energy deposition of particles. As is standard in 2PC analyses~\cite{HION-2015-09,HION-2016-01,HION-2018-27}, the correlation functions were then corrected for the impact of detector non-uniformity through a mixed-event procedure and normalized to a total integral of $2\pi$.
The resulting correlation functions, $Y(\Delta\phi,\eta^a)$, were expressed as a Fourier series, $Y = G \left\{ 1+ 2 \sum_{n=1}^{4} a_n(\eta^a) \cos(n\Delta\phi)\right\}$, where the $\eta^a$-dependent coefficients $a_n$ are the ``raw'' Fourier coefficients of order $n$ and $G$ is the normalization factor.
An example of the $a_n(\eta^a)$ values in \pp\ collisions is shown in Figure~\ref{fig:c2_exp}.

The contribution of non-flow to the correlation functions was removed using a template-based NFS method~\cite{HION-2015-09,HION-2019-03,HION-2018-27}. In this method, a low-multiplicity (LM) correlation function is used as a ``template'' that has a larger non-flow contribution, which then allows the non-flow contribution to be statistically removed from a high-multiplicity (HM) correlation function. To extract the flow coefficients, the template method relies on some assumptions, such as the Fourier $a_1$ term being completely due to non-flow and the shape of the non-flow correlation not changing with multiplicity.  The template method and associated assumptions are standard in flow measurement, and many have been validated in data
~\cite{HION-2013-04,HION-2015-09,Nagle:2021rep}.

In the 13~\TeV\ \pp\ data and the 5.02~\TeV\ \pp\ data (used only as the \xexe LM reference), the LM correlation function is constructed using events with multiplicity   $10\le\Nrec<30$. Each HM correlation function $Y^\mathrm{HM}(\Delta\phi)$ was fit with a weighted sum of the analogous LM correlation function, $Y^\mathrm{LM}(\Delta\phi)$, and an azimuthally modulated pedestal characterized by a set of flow coefficients $c_n$, i.e.,
$Y^\mathrm{HM}(\Delta\phi) = w_1Y^\mathrm{LM}(\Delta\phi) + w_2\left(1+2\sum_{n=2}^{4} c_n\cos(n\Delta\phi)\right)$,
where $w_{1,2}$ are the weight factors. This fitting was performed separately for each $\eta^a$ selection, resulting in $\eta^a$-dependent NFS coefficients $c_n(\eta^a)$, examples of which can also be seen in Figure~\ref{fig:c2_exp}.   Example 2PCs for individual $\eta^a$ bins and their associated template fits can be viewed in the Appendix.
 
The longitudinal decorrelation effect was measured by fitting the coefficients $c_n(\eta^a)$ and $a_n(\eta^a)$ to the quadratic functional form $A_n \left(1+ F_n \cdot \eta^a + S_n \cdot \left(\eta^{a}\right)^{2}\right)$, with the following three free parameters: $A_n$, which accounts for the overall magnitude of the coefficients; $F_n$, which describes the relative decrease of $c_n$ or $a_n$ with $\eta^a$ and thus characterizes the linear decorrelation strength; and $S_n$, which models any higher-order dependence on $\eta^a$. Higher order terms $(\eta^a)^n$, i.e., $n>2$, provide negligible contributions.
In previous measurements~\cite{HION-2019-06,CMS-HIN-14-012}, the decorrelation parameter $F_n$ was defined differently.  These definitions are all nearly identical in the limit of small $F_n$, but the  definition used in this Letter is more well-suited for large raw $F_n$ values before NFS, see the Appendix for details.

\Fig{\ref{fig:c2_exp}} shows an example of the $\eta^a$-dependent correlation coefficients, measured in high-multiplicity ($60\le\Nrec<200$) 13~\TeV\ \pp collisions. The raw Fourier coefficients $a_2$ and $a_3$, which are sensitive to contributions from non-flow effects, decrease in magnitude as the gap between the $a$ particles (tracks) and reference particles (clusters) increases. The correlation coefficient $c_2$, which has had the NFS procedure applied, shows a weaker variation with track-cluster pseudorapidity separation. The opposite is true for coefficient $c_3$, which has a stronger variation with separation after NFS.
This is a result of contributions from non-flow to $a_n$, which are of opposite sign for $a_2$ and $a_3$.
The $\eta^a$-dependent differences between the $c_n$ and $a_n$ values highlight the need to account for non-flow effects in the measurement. The results of applying the $\eta^a$-dependent fits, used to extract the values of $F_n$, are also shown and provide a good characterization of the data. Similar examples of correlation coefficients as a function of $\eta^a$ in \xexe collisions can be viewed in the Appendix.

In the track-tower measurement, since the energy in nearby towers is correlated, a bootstrapping method using pseudo-experiment sampling~\cite{ATLAS:2021jhn} was used to determine the statistical uncertainties.
For both the track-tower and track-topocluster measurements, multiple potential sources of systematic uncertainty were evaluated, with those having a significant impact on the measurement summarized here.
The uncertainty associated with the choice of low-multiplicity reference in the template method (nominally $10\le\Nrec<30$) was evaluated by using alternative, but partially overlapping, ranges in \Nrec.
The uncertainty in the mixed-event correction procedure was evaluated by changing the required similarity of the mixed events in \Nrec and the $z$-coordinate of the vertex. The sensitivity to the track selection was evaluated by varying the track quality criteria and the reconstruction efficiency $\epsilon^\mathrm{trk}$ by its uncertainty~\cite{STDM-2015-02}. The sensitivity to the trigger composition in \pp\ data was evaluated by raising the trigger efficiency requirement of 70\% to 80\%. Different uncertainty sources were dominant in different \Nrec ranges, collision systems, and observables, with no single one dominant. For the \pp\ results, the statistical uncertainties are larger than systematic uncertainties in all cases.
 
\begin{figure}[t]
\centering
\includegraphics[width=0.49\textwidth]{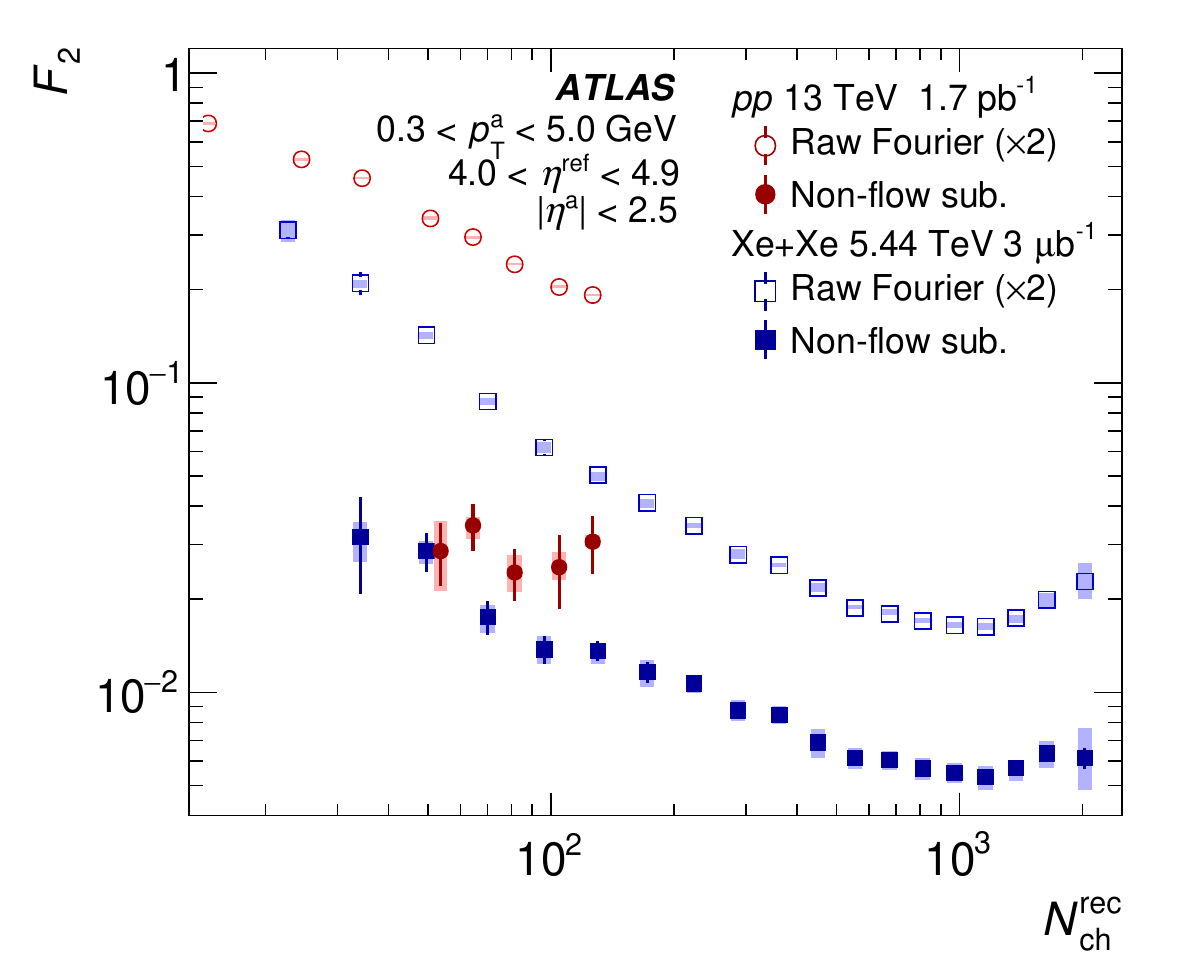}
\includegraphics[width=0.49\textwidth]{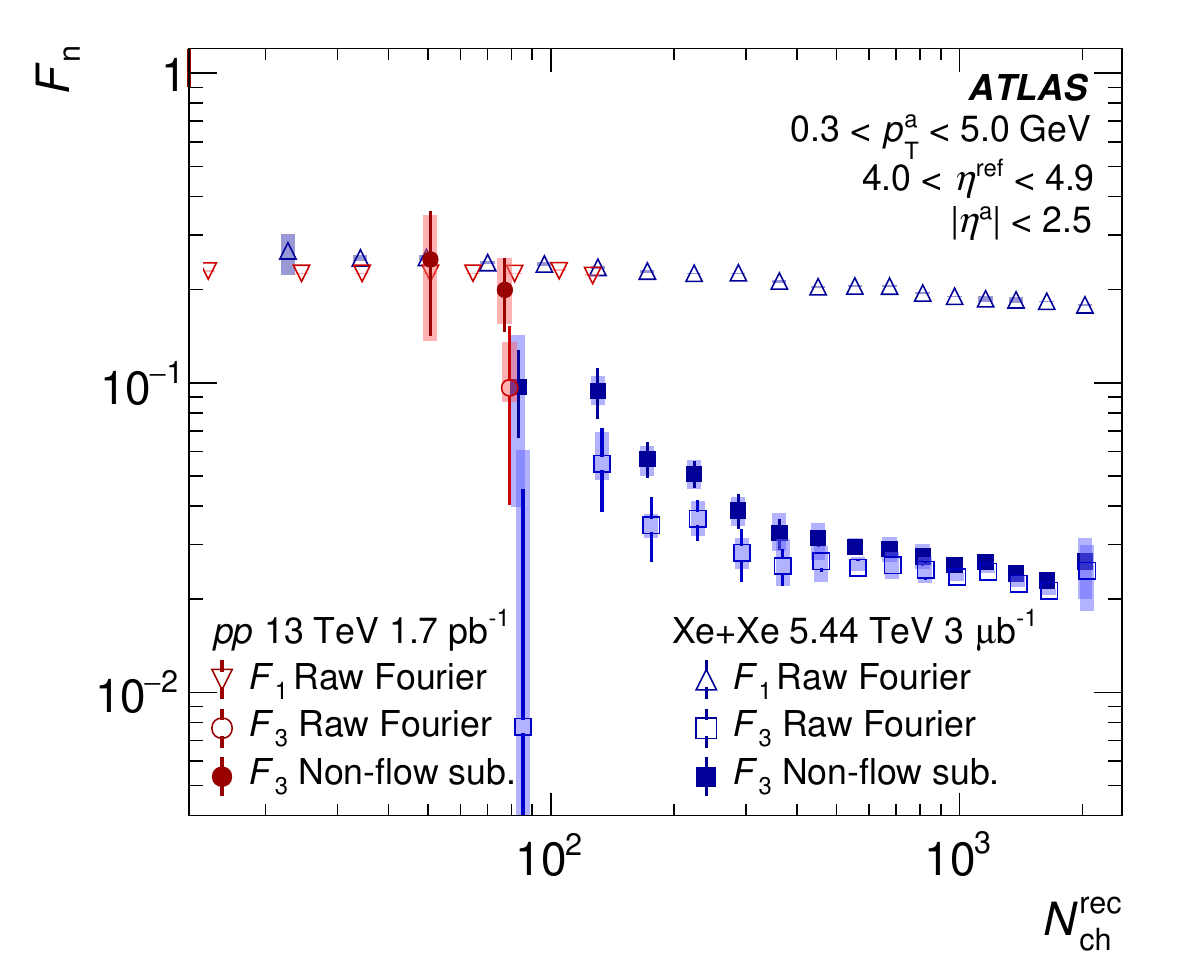}
\caption{Longitudinal decorrelation parameters $F_n$ for 13~\TeV\ \pp (red) and 5.44~\TeV\ \xexe (blue) for $n=2$ (left) and $n=1,3$ (right) as a function of \Nrec. The parameters are reported after the application of the NFS to the correlation function (solid markers), and those derived from fits to the raw Fourier coefficients (open markers).
The statistical (systematic) uncertainties are drawn as vertical lines (shaded boxes).  Points with errors exceeding 0.15 are not shown.
}
\label{fig:results_comb}
\end{figure}

\Fig{\ref{fig:results_comb}} presents the longitudinal decorrelation parameters $F_n$ in \pp collisions at 13~\TeV\ and \xexe collisions at 5.44~\TeV\ as a function of \Nrec. The measured $F_n$ values extracted from fits to the correlation coefficients $c_n(\eta^a)$ are shown for $n=2,3$, as well as raw values for $n=1,2,3$ obtained from quadratic fits to the Fourier coefficients $a_n(\eta^a)$ before the application of the NFS.
 
In \pp collisions, the raw $F_2$ values decrease strongly with increasing \Nrec, reflecting the changing relative contribution from flow and non-flow. With NFS, an \Nrec-independent $F_2 \approx 0.02$--$0.03$ is observed.
The data also yields a measurement of NFS $F_3$, which is large and positive at $\approx0.2$.
Notably, the effect of NFS is to decrease the $F_2$ values while increasing those of $F_3$.
 
As in \pp, the raw $F_2$ in \xexe decreases strongly with \Nrec, but with a small increase in the highest \Nrec events. These results are similar to the previous measurement performed without NFS~\cite{HION-2019-06}. The NFS $F_2$ values are smaller than the raw ones by more than a factor of three (two) at low \Nrec (large \Nrec), and range from $\approx0.03$ to $\approx0.005$ (note the $\times2$ applied to the raw Fourier in the left panel of Fig.~\ref{fig:results_comb}). The measurement of $F_3$ is less sensitive to the application of NFS, and the resulting values are larger than those for $F_2$ over most of the multiplicity range.
 
Even in the highest \Nrec \xexe collisions, there are important non-flow corrections to $F_2$ and $F_3$.     However, the template fitting method assumes that there is no modification to the non-flow correlation shape as a function of \Nrec.
Jet quenching effects, where partons lose energy in the QGP, are known to suppress the yields of jet-correlated particles and modify their angular distributions~\cite{Connors:2017ptx,ALICE:2011gpa}. Additionally, the slow change of $F_1$ with \Nrec, which reflects only non-flow, may indicate a shape modification or the presence of another source of first harmonic correlations~\cite{HION-2013-04,Retinskaya:2012ky}. Further exploration of these possible effects, or other observables less sensitive to non-flow~\cite{Xu:2020koy}, may be required for precision data-model comparisons in this large-\Nrec region.
 
For the template fit to give an unbiased measurement of flow decorrelation, the flow decorrelation in the template and the HM data must be similar, see the Appendix for details.   In the case of the \pp analysis, the NFS $F_2$ results are consistent with an \Nrec-independent flow decorrelation and thus no bias is incurred.
However, the \xexe $F_2$ is not \Nrec independent, thus a precise calculation of the possible incurred bias has to be performed.  Given a range of possible 5.02~\TeV\ \pp $F_2$ values of 0.00--0.07, this potential bias was determined to be at the 0 to +10\% level.
 
At the lowest multiplicities, $\Nrec \approx 30$--$50$, where the \xexe events have a large contribution from single nucleon-nucleon collision configurations, the $F_2$ values with NFS are similar to those in \pp events.  At larger multiplicities, the $F_2$ values remain large in \pp\ collisions but decrease in \xexe collisions, indicating that the
mechanism of additional particle production
is different in the two systems. These results, along with the lack of \Nrec dependence of azimuthal anisotropy in \pp collisions~\cite{HION-2016-01}, suggest that the correlation between the initial-state geometry and overall particle production is different at sub-nucleonic scales than at nucleonic scales.  The $F_3$ results in \pp\ and \xexe collisions are statistically compatible, agreeing with the qualitative picture indicated by the $F_2$ results.
 
\begin{figure}
\centering
\includegraphics[width=0.49\textwidth]{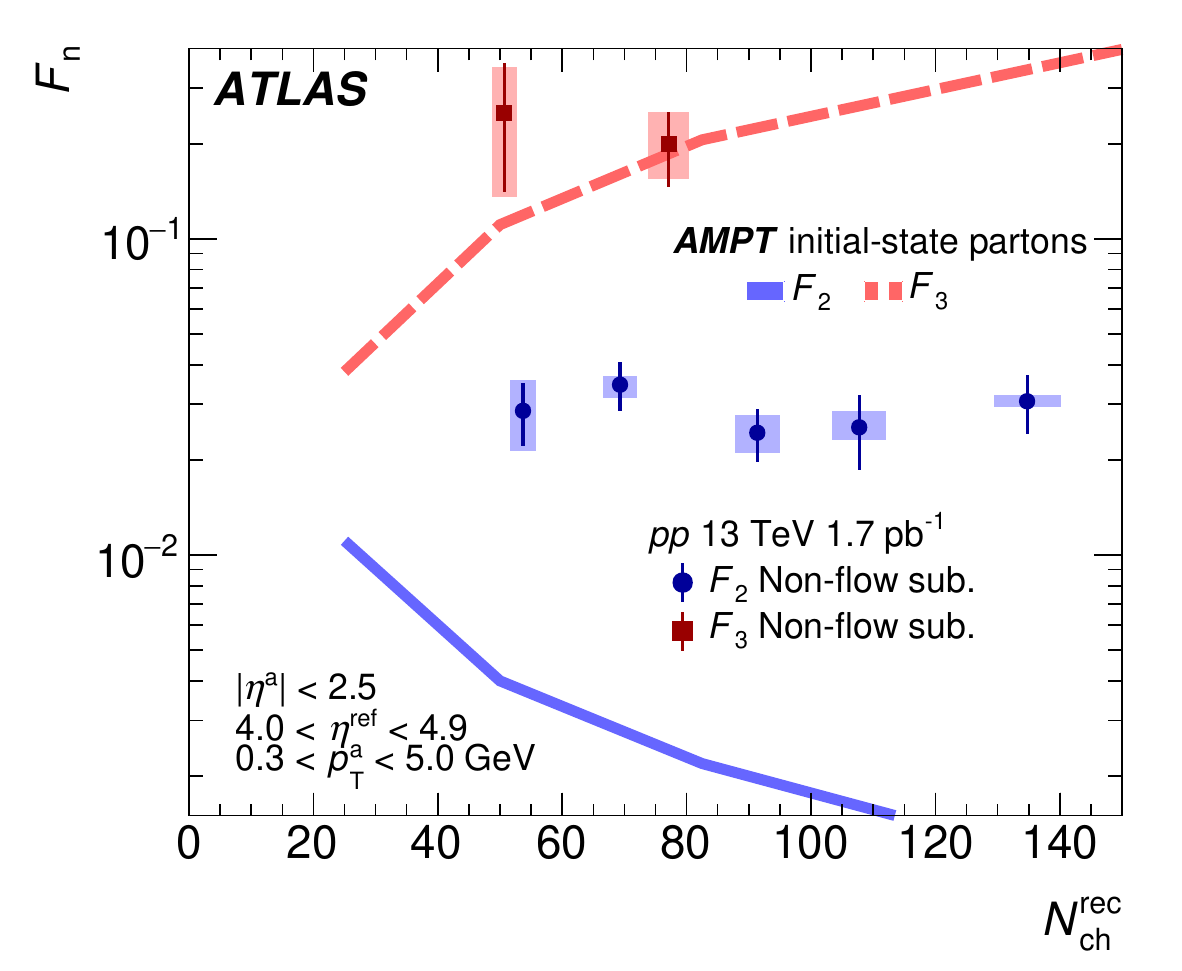}
\includegraphics[width=0.49\textwidth]{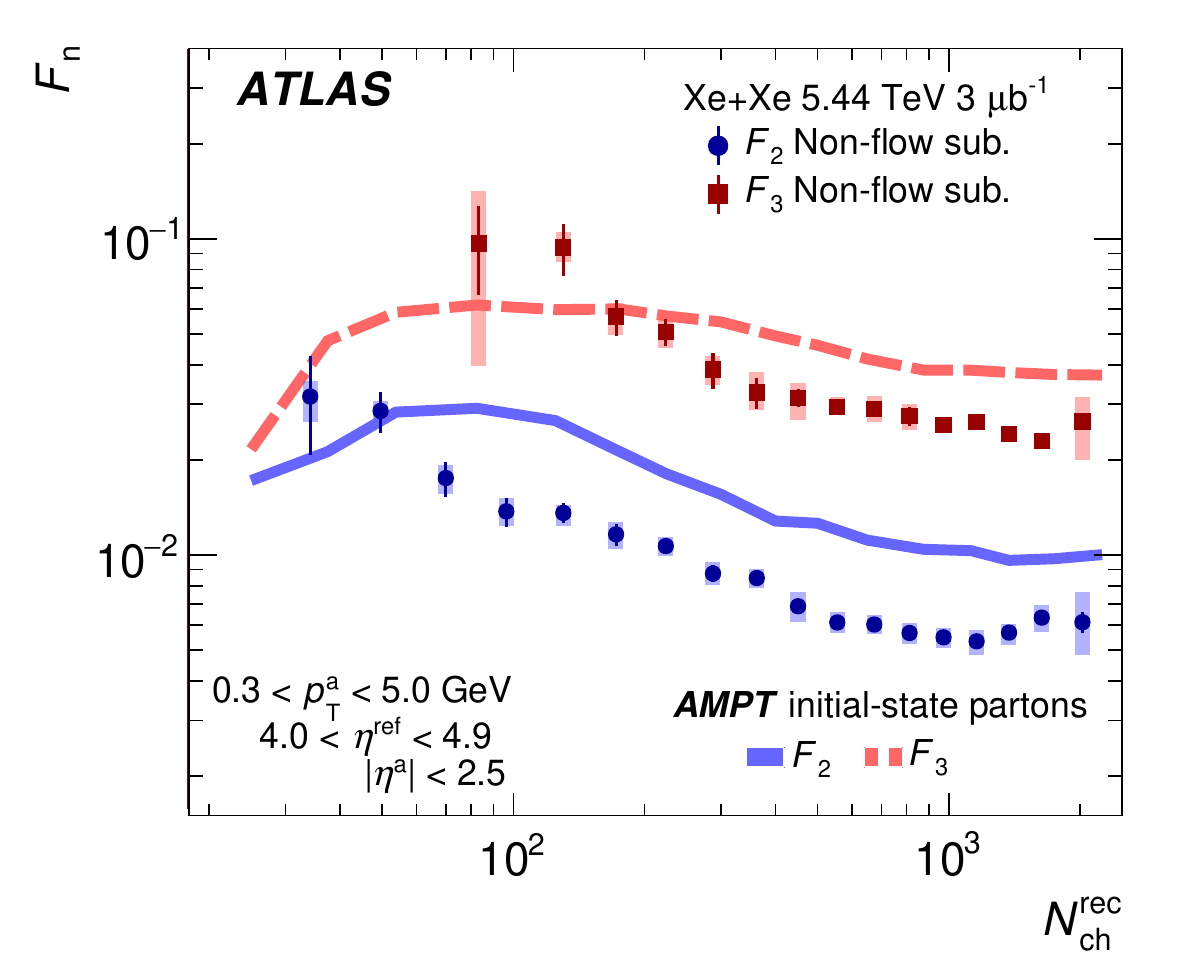}
\caption{Comparison of \textsc{AMPT} theory  calculations to the 13~\TeV\ \pp (left) and  5.44~\TeV\ \xexe (right) results.  In data, the non-flow subtracted $F_n$ values are shown as a function of \Nrec.  The statistical (systematic) uncertainties are drawn as vertical lines (shaded bands).
}
\label{fig:ampt}
\end{figure}
 
The data are further compared with calculations from the \textsc{AMPT} model~\cite{Lin:2004en} as shown in Fig.~\ref{fig:ampt}.  \textsc{AMPT} is a kinetic transport model that starts from a full three-dimensional initial state of color strings, evolves partons from those strings, the partons scatter, and finally hadronize via a coalescence prescription.  \textsc{AMPT} has been shown to reproduce a broad range of observables related to bulk collectivity in nucleus-nucleus collisions~\cite{Lin:2021mdn}.
Since the initial-state geometry in \textsc{AMPT} is known, the $F_n$ parameters can be calculated directly via the decorrelation of the spatial eccentricity vectors~\cite{Pang:2015zrq} as a function of pseudorapidity separation, thus avoiding the non-flow effects introduced by 2PC methods.
The spatial eccentricity vectors are calculated from the positions of the partons when they form via the color strings.
The \textsc{AMPT} \xexe charged particle multiplicity was scaled by a factor of 0.5, to reproduce the distribution of \Nrec in data.  No rescaling was applied to the \textsc{AMPT} \pp predictions.
 
For \pp collisions, the simple geometric description in \textsc{AMPT} in terms of two color strings spanning a large longitudinal extent (thus resulting in small $F_2$) is highly disfavored by the data.   In  contrast the $F_3$ is well described and should arise primarily via geometry fluctuations.
In the case of \xexe, \textsc{AMPT} shows a qualitative agreement for the multiplicity dependence of $F_{2,3}$, though a factor of two higher than the data for \Nrec $> 250$.
Moving to lower \Nrec \xexe collisions, the \textsc{AMPT} $F_{2,3}$ values turn over and begin to decrease, which is not present in the data. This decreasing behavior is likely due to the prevalence of geometries with very few strings, including the single nucleon-nucleon (i.e., \pp-like) case.
 
In conclusion, this paper presents the measurements of longitudinal flow decorrelation parameters $F_n$ in 13~\TeV \pp and 5.44~\TeV \xexe collisions. These results are derived from pseudorapidity-separated two-particle correlations, after the application of a  template-based non-flow subtraction. Both the systems feature large decorrelation signals in low-multiplicity events that are \Nrec-independent in \pp and decrease with increasing \Nrec in \xexe.
The data were compared with the \textsc{AMPT} color-string geometry calculations, which have a good qualitative modeling for the large \Nrec \xexe data but significantly under-predict the decorrelation effects in single nucleon-nucleon collisions, both in the \pp collisions and potentially the \xexe collisions.
The results when compared with a color string-based model of the initial geometry indicate that in \pp and low-multiplicity \xexe collisions, sub-nucleonic structure and fluctuations in longitudinal energy deposition are needed to describe the data.   In contrast, high-multiplicity \xexe and Pb+Pb collisions have longitudinal decorrelations reasonably understood in terms of intrinsic nuclear geometry and fluctuating longitudinal energy deposition at the nucleon scale.
 
\section*{Acknowledgments}


We thank CERN for the very successful operation of the LHC, as well as the
support staff from our institutions without whom ATLAS could not be
operated efficiently.
 
We acknowledge the support of
ANPCyT, Argentina;
YerPhI, Armenia;
ARC, Australia;
BMWFW and FWF, Austria;
ANAS, Azerbaijan;
CNPq and FAPESP, Brazil;
NSERC, NRC and CFI, Canada;
CERN;
ANID, Chile;
CAS, MOST and NSFC, China;
Minciencias, Colombia;
MEYS CR, Czech Republic;
DNRF and DNSRC, Denmark;
IN2P3-CNRS and CEA-DRF/IRFU, France;
SRNSFG, Georgia;
BMBF, HGF and MPG, Germany;
GSRI, Greece;
RGC and Hong Kong SAR, China;
ISF and Benoziyo Center, Israel;
INFN, Italy;
MEXT and JSPS, Japan;
CNRST, Morocco;
NWO, Netherlands;
RCN, Norway;
MEiN, Poland;
FCT, Portugal;
MNE/IFA, Romania;
MESTD, Serbia;
MSSR, Slovakia;
ARRS and MIZ\v{S}, Slovenia;
DSI/NRF, South Africa;
MICINN, Spain;
SRC and Wallenberg Foundation, Sweden;
SERI, SNSF and Cantons of Bern and Geneva, Switzerland;
MOST, Taiwan;
TENMAK, T\"urkiye;
STFC, United Kingdom;
DOE and NSF, United States of America.
In addition, individual groups and members have received support from
BCKDF, CANARIE, Compute Canada and CRC, Canada;
PRIMUS 21/SCI/017 and UNCE SCI/013, Czech Republic;
COST, ERC, ERDF, Horizon 2020 and Marie Sk{\l}odowska-Curie Actions, European Union;
Investissements d'Avenir Labex, Investissements d'Avenir Idex and ANR, France;
DFG and AvH Foundation, Germany;
Herakleitos, Thales and Aristeia programmes co-financed by EU-ESF and the Greek NSRF, Greece;
BSF-NSF and MINERVA, Israel;
Norwegian Financial Mechanism 2014-2021, Norway;
NCN and NAWA, Poland;
La Caixa Banking Foundation, CERCA Programme Generalitat de Catalunya and PROMETEO and GenT Programmes Generalitat Valenciana, Spain;
G\"{o}ran Gustafssons Stiftelse, Sweden;
The Royal Society and Leverhulme Trust, United Kingdom.
 
The crucial computing support from all WLCG partners is acknowledged gratefully, in particular from CERN, the ATLAS Tier-1 facilities at TRIUMF (Canada), NDGF (Denmark, Norway, Sweden), CC-IN2P3 (France), KIT/GridKA (Germany), INFN-CNAF (Italy), NL-T1 (Netherlands), PIC (Spain), ASGC (Taiwan), RAL (UK) and BNL (USA), the Tier-2 facilities worldwide and large non-WLCG resource providers. Major contributors of computing resources are listed in Ref.~\cite{ATL-SOFT-PUB-2021-003}.


\section*{Appendix}
 
\subsection*{Correlation functions and coefficients}
 
\begin{figure}[hb!]
\centering
\includegraphics[width=0.24\textwidth]{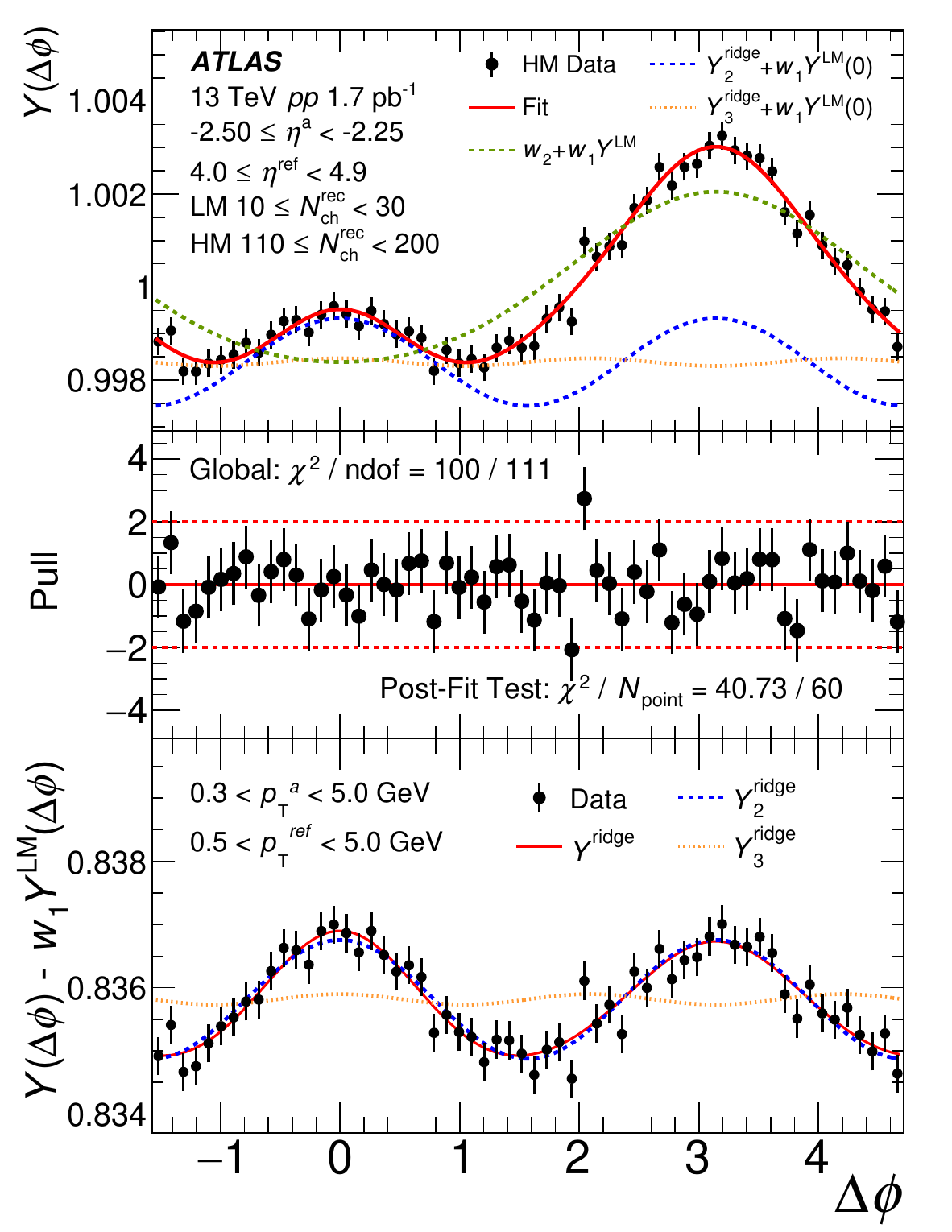}
\includegraphics[width=0.24\textwidth]{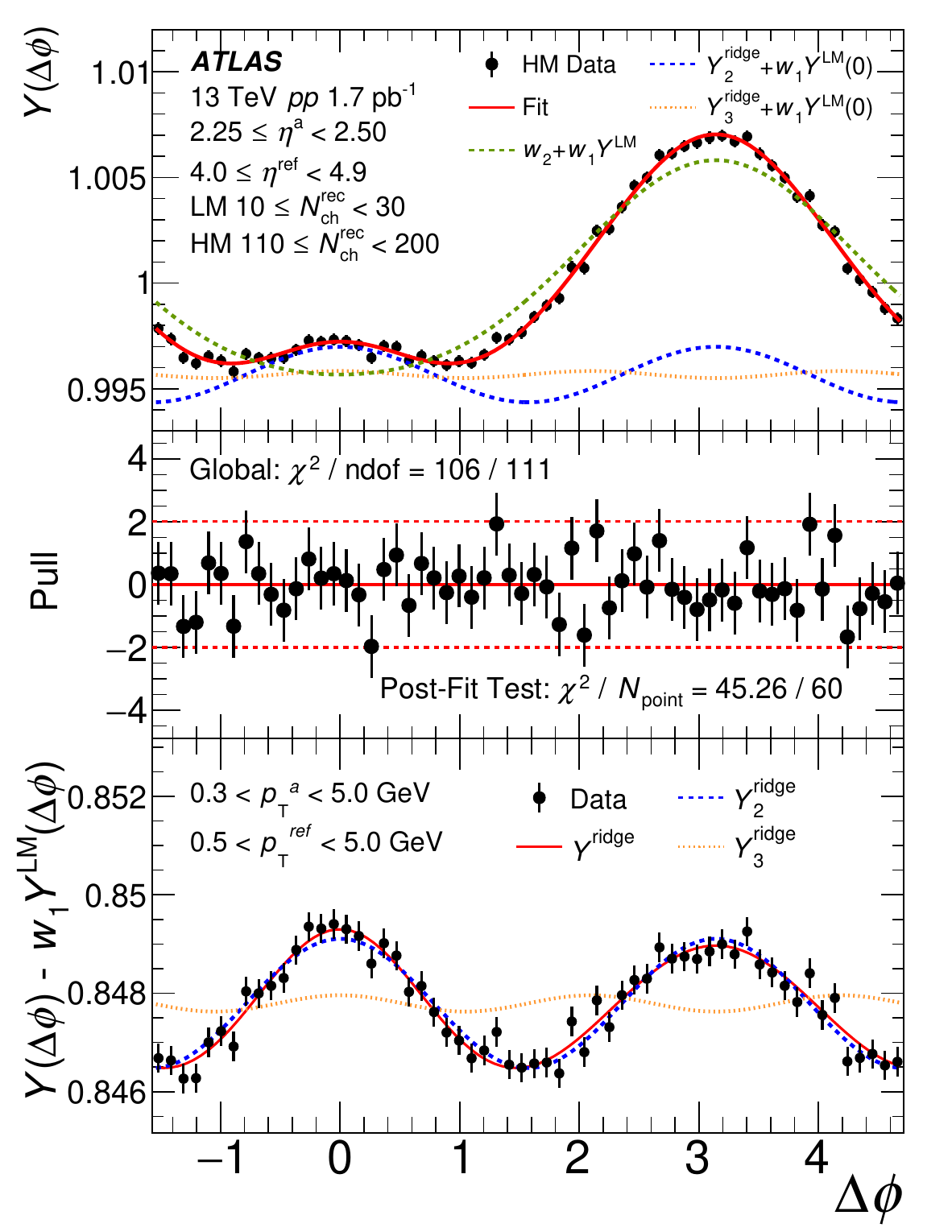}
\includegraphics[width=0.24\textwidth]{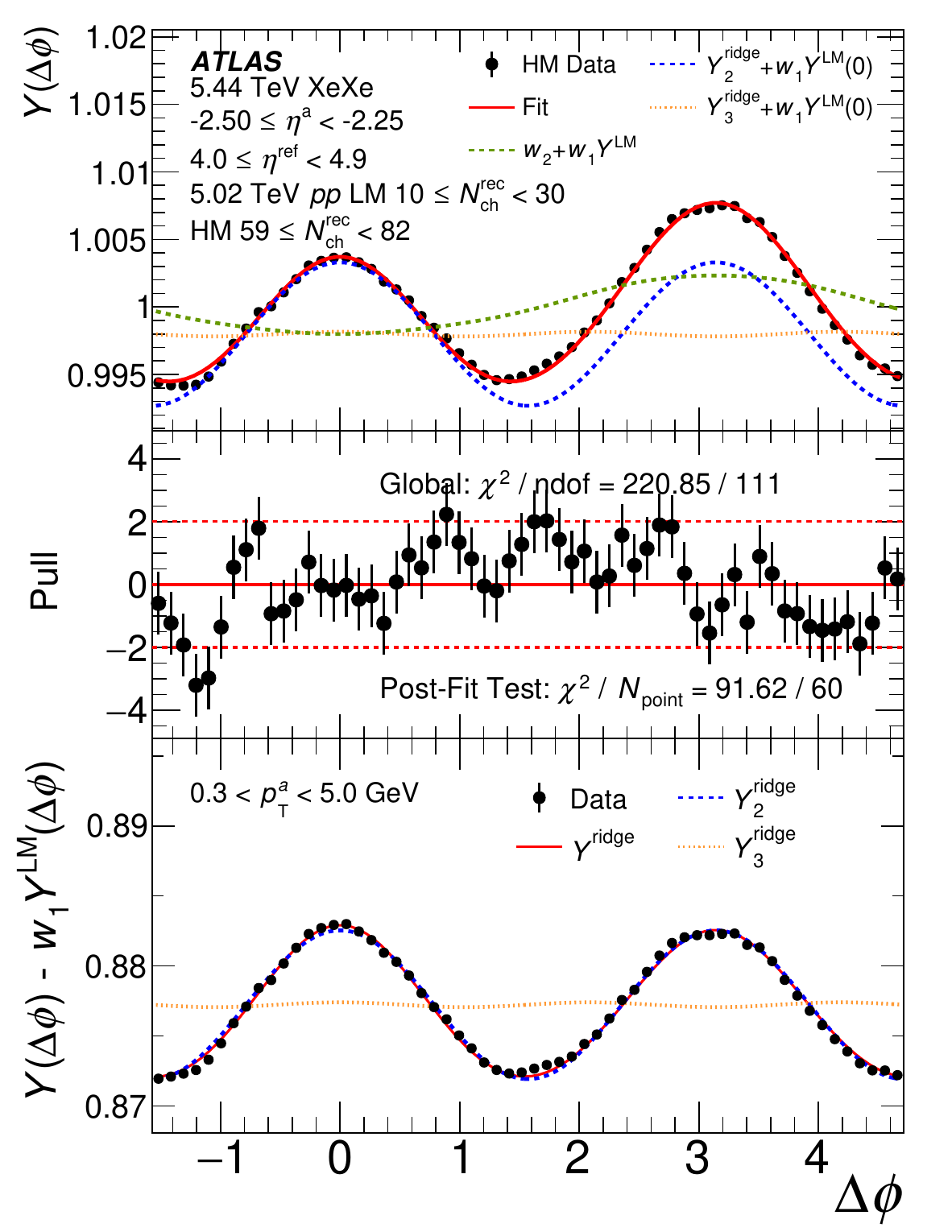}
\includegraphics[width=0.24\textwidth]{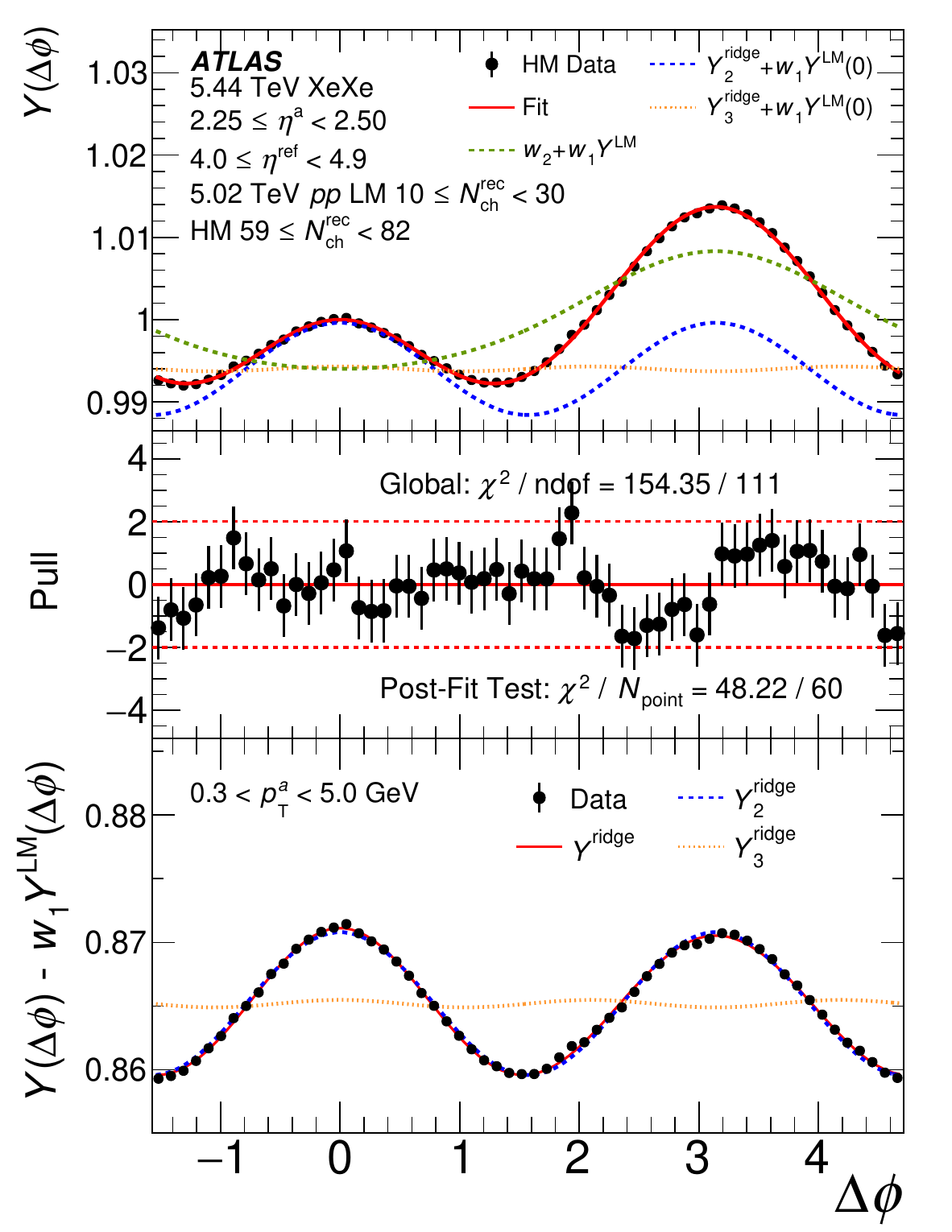}
\caption{Selected example 2PC template fit results for two $\eta^a$ intervals in 13~\TeV \pp collisions (left) and 5.44~\TeV \xexe collisions (right). In the top panels, the solid red line shows the total fit to the HM data, shown in black markers. The dashed green line shows the scaled LM reference correlation plus pedestal, while the dashed blue and dotted yellow lines indicate the two flow contributions to the fit, $Y_2^\mathrm{ridge} = w_2[1 +  2c_{2}\cos(2\Delta\phi)]$ and $Y_3^\mathrm{ridge} = w_2 [1 + 2c_{3}\cos(3\Delta\phi)]$, shifted upwards by $w_1 Y^\mathrm{LM}(0)$ for visibility.
The middle panels show the pull distribution for the template fits.
The global $\chi^2$/ndf is calculated from the simultaneous fit to the HM and LM correlations. The post-fit $\chi^2$/$N_{\mathrm{point}}$ is calculated from the pull distribution in the panel.
The bottom panels show the same set of data and fit components, where the scaled LM distribution was subtracted from each to better isolate the modulation.}
\label{fig:tempFit}
\end{figure}

Examples of the 2PC functions are shown in Fig.~\ref{fig:tempFit} for \pp in the left two panels and Xe+Xe in the right two panels for different \Nrec selections.
In all cases, the ID charged-particle tracks are used as $a$-type particles and FCal topoclusters (towers) are used as reference particles for  \pp collisions (Xe+Xe collisions).  In all cases, there is peak-like feature centered at $\Delta\Phi=\pi$, which is dominated by contributions from non-flow correlations, specifically dijets, and is well  described by the low multiplicity template $Y^\text{LM}(\Delta\Phi)$. The residual flow-like correlation after NFS is shown in the lower panels.
 
\begin{figure}
\centering
\includegraphics[width=0.4\textwidth]{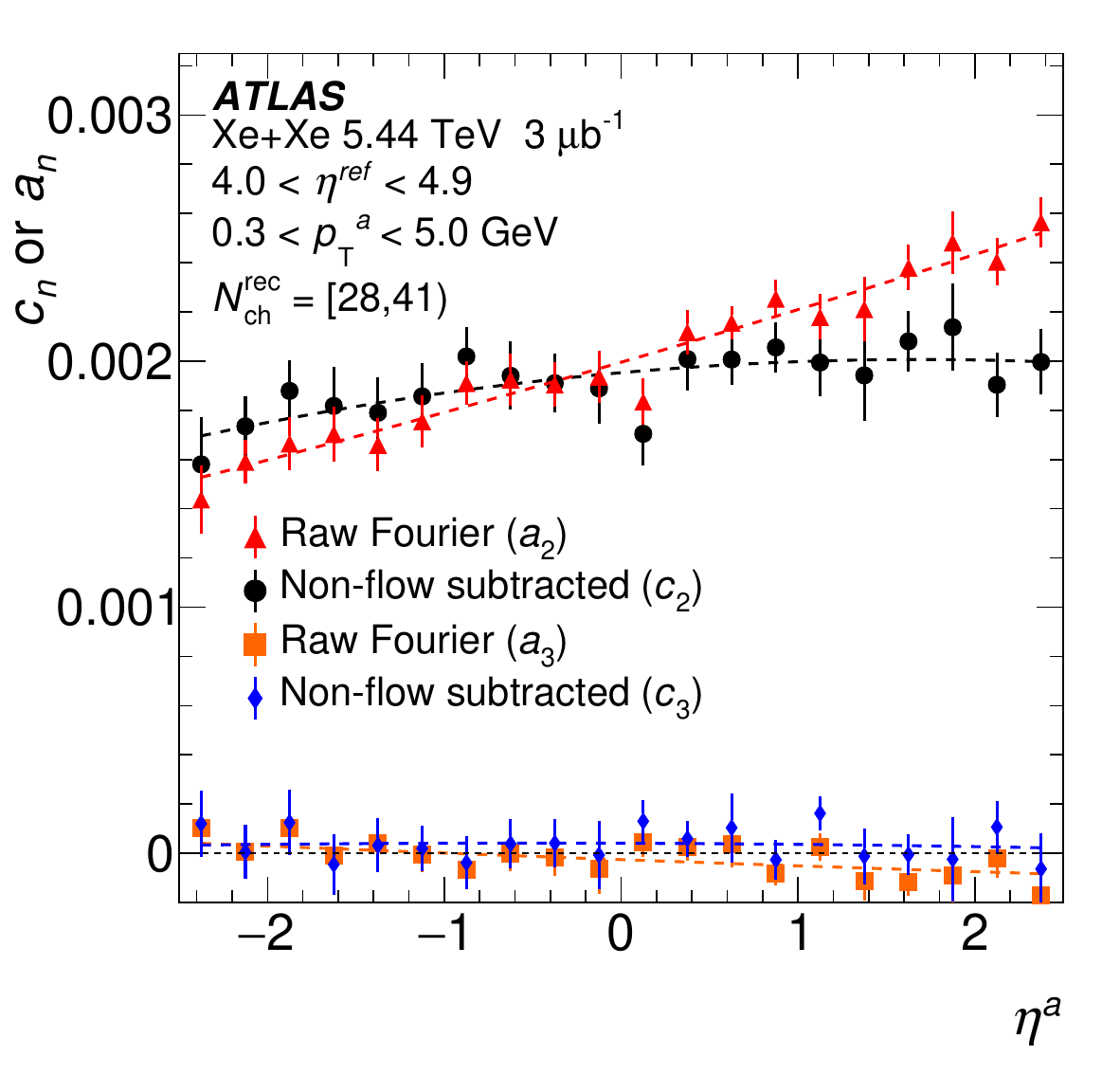}
\includegraphics[width=0.4\textwidth]{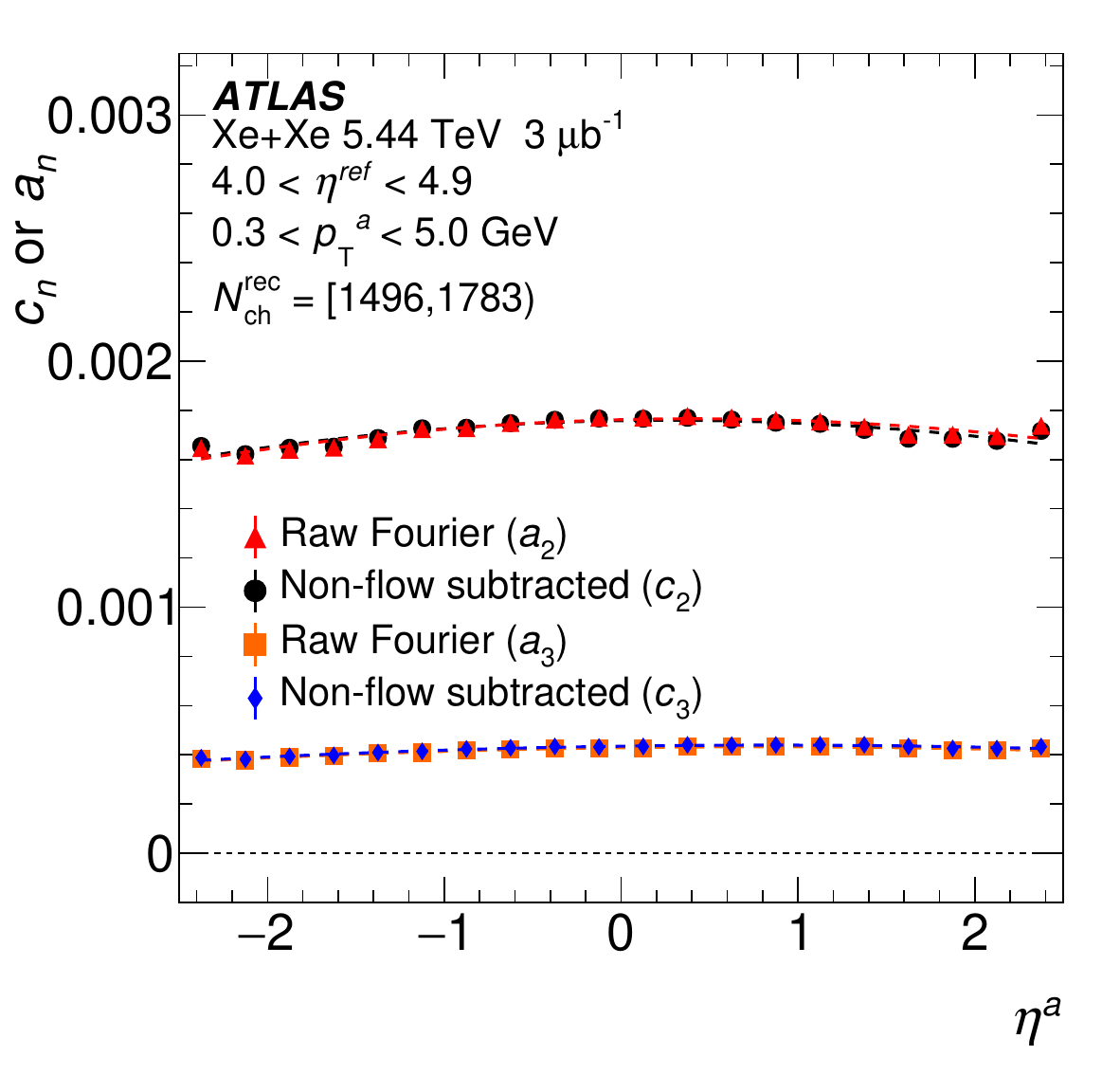}
\caption{Coefficients $a_n$ and $c_n$ in low-multiplicity (high-multiplicity) 5.44~\TeV\ \xexe events are shown in the left (right) panel, for $n = 2, 3$ as a function of track pseudorapidity $\eta^a$.  The $c_n$ values are obtained from the template fitting method.  The dashed lines represent a fit to a quadratic function in $\eta^a$ for each set of coefficients.
The pseudorapidity gap between the tower and track pair is largest for negative $\eta^a$ and decreases moving to positive $\eta^a$. Only the statistical uncertainties are shown.  }
\label{fig:c2_exp_xexe}
\end{figure}

The resulting correlation coefficients from the raw Fourier fit and NFS for \xexe collisions in two \Nrec ranges are shown in Fig.~\ref{fig:c2_exp_xexe}.
 
\clearpage
 
\subsection*{Further details on decorrelation extraction}
 
There are two additional details that are important for (i) relating these results to other experimental measurements of longitudinal flow decorrelations and for (ii) understanding potential biases introduced in the template NFS procedure from one of nultiple assumptions.
 
First, in previous heavy-ion analyses of longitudinal decorrelations~\cite{HION-2016-04,HION-2019-06}, a set of specific ratios of correlation coefficients were taken.  The $r_n$ ratio as a function of \etaA, is defined as
\begin{equation}
r_n(\etaA) =  \frac{c_{n}(-\etaA)}{c_{n}(\etaA)},
\end{equation}
which also applies to coefficient $a_n$ and can be substituted for $c_n$.   If $c_n$ is thought of as $c_n(\eta^a) \propto 1+F_n\eta^a$, which is the case in this paper, then the above equation yields $r_n(|\eta^a|) \approx 1-2F_n|\eta^a| + 2F_n^2|\eta^a|^2$.
Thus a fit to $r_n(\etaA)$ with $1-2P_0\etaA$ can be used to extract $F_n$, where $F_n=P_0$.  However, if $F_n$ is large, the next higher-order term in \etaA leads to a systematically smaller extracted $F_n$.  The decorrelations measured in large systems in the literature, as well as the coefficients with NFS in this analysis, are sufficiently small that this effect is negligible.   In contrast the measured raw Fourier moments in small systems have large non-linear trends with $\eta^a$ and such a method leads to underestimates of $F_n$.
 
Second, the true flow coefficients of the HM correlation, $c_n^\text{HM}(\eta^a)$, are not measured directly by the template-based NFS.  Rather, the $c_n(\eta^a)$ coefficients are measured and a set of criteria need to be met for these measured coefficients to be exactly  equal to $c_n^\text{HM}(\eta^a)$.
Similarly, when measuring the linear moment of the $\eta^a$ dependence of $c_n(\eta^a)$ through the methods detailed in this Letter, the resulting, $F_n$, has a set of criteria that must be fulfilled for this measured flow decorrelation to correspond to the true flow decorrelation in the HM selection $F_n^\text{HM}$.
These criteria relate to the mid-rapidity flow $c_n^\text{LM}|^{\eta^a=0}$ and flow decorrelation $F_{n}^\text{LM}$ in the template LM selection.  The mathematical representation of the criteria is
\begin{equation}
F_n \approx \frac{c_{n}^\mathrm{HM}|^{\eta^a=0} F_n^\mathrm{HM}-w_1 c_n^\mathrm{LM}|^{\eta^a=0}F_n^\mathrm{LM}}{c_{n}^\mathrm{HM}|^{\eta^a=0}-w_1 c_{n}^\mathrm{LM}|^{\eta^a=0}} .
\label{eq:crtdTemp_eq}
\end{equation}
One observes implicitly that if $F_n^\text{LM}$ and $F_n^\text{HM}$ are equal, $F_n$ corresponds to the true flow decorrelation in the HM selection. Within uncertainties, this is the case for the $F_n$ results in 13~\TeV \pp collisions.  For the \xexe results, because  the extracted $F_n$ values are found to vary with \Nrec, the above equation can be used to measure the potential bias.  Inserting all the parameters and varying $F_n^{LM}$ from 0.0--0.07, the deviation of $F_n^\text{HM}$ and the reported $F_n$ is found to be 0 to 10\% level.
Because $F_n^\text{LM}$ is the true flow decorrelation (after NFS), with the current technique, it is impossible to know what it is because this multiplicity range is used for the NFS itself by definition.  Thus the range used to quantify this specific possible bias in the template fitting procedure (0 to 0.07) is arbitrarily chosen to demonstrate that even a large range of possible $F_n^\text{LM}$ produces systematic effects that are about the size of the uncertainties of the measurement.
 
Eq.~\ref{eq:crtdTemp_eq} can be derived by starting with Eq.~14 from Ref.~\cite{Lim:2019cys}, introducing a linear dependence of the correlation coefficients on $\eta^a$, constructing the $r_n$ ratio from Eq.~\ref{eq:crtdTemp_eq}, and setting it equal to $1-2F_n$.

\printbibliography
 
\clearpage
 
\begin{flushleft}
\hypersetup{urlcolor=black}
{\Large The ATLAS Collaboration}

\bigskip

\AtlasOrcid[0000-0002-6665-4934]{G.~Aad}$^\textrm{\scriptsize 102}$,
\AtlasOrcid[0000-0002-5888-2734]{B.~Abbott}$^\textrm{\scriptsize 120}$,
\AtlasOrcid[0000-0002-1002-1652]{K.~Abeling}$^\textrm{\scriptsize 55}$,
\AtlasOrcid[0000-0001-5763-2760]{N.J.~Abicht}$^\textrm{\scriptsize 49}$,
\AtlasOrcid[0000-0002-8496-9294]{S.H.~Abidi}$^\textrm{\scriptsize 29}$,
\AtlasOrcid[0000-0002-9987-2292]{A.~Aboulhorma}$^\textrm{\scriptsize 35e}$,
\AtlasOrcid[0000-0001-5329-6640]{H.~Abramowicz}$^\textrm{\scriptsize 151}$,
\AtlasOrcid[0000-0002-1599-2896]{H.~Abreu}$^\textrm{\scriptsize 150}$,
\AtlasOrcid[0000-0003-0403-3697]{Y.~Abulaiti}$^\textrm{\scriptsize 117}$,
\AtlasOrcid[0000-0002-8588-9157]{B.S.~Acharya}$^\textrm{\scriptsize 69a,69b,q}$
% ...
\bigskip
\\

% ...
$^{29}$Physics Department, Brookhaven National Laboratory, Upton NY; United States of America.\\
% ...
$^{35}$$^{(a)}$Facult\'e des Sciences Ain Chock, R\'eseau Universitaire de Physique des Hautes Energies - Universit\'e Hassan II, Casablanca;$^{(b)}$Facult\'{e} des Sciences, Universit\'{e} Ibn-Tofail, K\'{e}nitra;$^{(c)}$Facult\'e des Sciences Semlalia, Universit\'e Cadi Ayyad, LPHEA-Marrakech;$^{(d)}$LPMR, Facult\'e des Sciences, Universit\'e Mohamed Premier, Oujda;$^{(e)}$Facult\'e des sciences, Universit\'e Mohammed V, Rabat;$^{(f)}$Institute of Applied Physics, Mohammed VI Polytechnic University, Ben Guerir; Morocco.\\
% ...
$^{49}$Fakult\"{a}t Physik , Technische Universit{\"a}t Dortmund, Dortmund; Germany.\\
% ...
$^{55}$II. Physikalisches Institut, Georg-August-Universit\"{a}t G\"ottingen, G\"ottingen; Germany.\\
% ...
$^{69}$$^{(a)}$INFN Gruppo Collegato di Udine, Sezione di Trieste, Udine;$^{(b)}$ICTP, Trieste;$^{(c)}$Dipartimento Politecnico di Ingegneria e Architettura, Universit\`a di Udine, Udine; Italy.\\
% ...
$^{102}$CPPM, Aix-Marseille Universit\'e, CNRS/IN2P3, Marseille; France.\\
% ...
$^{117}$Department of Physics, New York University, New York NY; United States of America.\\
% ...
$^{120}$Homer L. Dodge Department of Physics and Astronomy, University of Oklahoma, Norman OK; United States of America.\\
% ...
$^{150}$Department of Physics, Technion, Israel Institute of Technology, Haifa; Israel.\\
$^{151}$Raymond and Beverly Sackler School of Physics and Astronomy, Tel Aviv University, Tel Aviv; Israel.\\
% ...
$^{a}$ Also Affiliated with an institute covered by a cooperation agreement with CERN.\\
$^{b}$ Also at An-Najah National University, Nablus; Palestine.\\
$^{c}$ Also at APC, Universit\'e Paris Cit\'e, CNRS/IN2P3, Paris; France.\\
$^{d}$ Also at Borough of Manhattan Community College, City University of New York, New York NY; United States of America.\\
$^{e}$ Also at Center for High Energy Physics, Peking University; China.\\
$^{f}$ Also at Center for Interdisciplinary Research and Innovation (CIRI-AUTH), Thessaloniki ; Greece.\\
% ...
$^{q}$ Also at Department of Physics, King's College London, London; United Kingdom.\\
% ...

\end{flushleft}


\end{document}